\documentclass[12pt,a4paper]{article}
\usepackage[T2A]{fontenc}
\usepackage[utf8]{inputenc}
\usepackage[english]{babel}
\usepackage{amssymb}
\usepackage{amsmath}
\usepackage{graphicx}
\usepackage[small,sc,center]{titlesec}
\usepackage{indentfirst}
\usepackage{siunitx}
\usepackage[labelsep=period]{caption}
\usepackage{caption}
\newcommand{\nc}{\newcommand}
\nc{\etab}{\eta_\mathrm{B}}
\nc{\teff}{T_\mathrm{eff}}
\nc{\tev}{t_\mathrm{ev}}
\begin{document}
\begin{center}
	\textbf{Radial pulsations of stars on the stage of the final helium flash}
	
	\vskip 3mm
	\textbf{Yu. A. Fadeyev\footnote{E--mail: fadeyev@inasan.ru}}
	
	\textit{Institute of Astronomy, Russian Academy of Sciences,
		Pyatnitskaya ul. 48, Moscow, 119017 Russia} \\
	
	Received August 12, 2019; revised August 13, 2019; accepted August 13, 2019
\end{center}

\textbf{Abstract} ---
Stellar evolution calculations to the stage of the cooling white dwarf were done for
population~I stars with masses on the main sequence $1M_\odot\le M_0\le 1.5M_\odot$.
The final helium flash LTP is shown to occur in post--AGB stars with initial masses
$1.3M_\odot\le M_0\le 1.32M_\odot$ for the overshooting parameter $f=0.016$.
In the case of more effective overshooting ($f=0.018$) the final helium flash occurs
at initial masses $1.28M_\odot\le M_0\le 1.3M_\odot$.
Fivefold variations of the parameter responsible for the mass loss rate during the
post--AGB stage do not affect occurrence of the final helium flash but lead to perceptible
changes of the evolutionary time.
Selected models of two evolutionary sequences with initial mass $M_0 = 1.3M_\odot$
computed with overshooting parameters $f=0.016$ and $f=0.018$ were used as initial
conditions in solution of the equations of hydrodynamics describing radial oscillations
of stars on the stage of the final helium flash at effective temperatures $\teff < 10^4$\:K.
The maximum pulsation period $\Pi=117$ day determined for the evolutionary sequence
$M_0=1.3M_\odot$, $f=0.016$ is in a good agreement with observational estimates of
the period of FG~Sge.
The mass, the radius and the effective temperature of the star are
$M=0.565M_\odot$, $R=126R_\odot$ and $\teff=4445$\:K, respectively.
At the same time the average period change rate of FG~Sge from 1960 to 1990 is nearly
three time larger than its theoretical estimate.


\newpage
\section*{introduction}

The end of the asymptotic giant branch (AGB) evolutionary stage of stars with
initial masses $M_0 \lesssim 9M_\odot$ is due to mass loss caused by the strong
stellar wind.
Decrease of the hydrogen envelope mass below $\sim 1\%$ of the stellar mass leads to
rapid contraction of the star so that in the Hertzsprung--Russel (HR) diagram the star
moves to the high--temperature domain with effective temperatures of $\teff\sim 10^5$~K
(Paczy\'nski 1971).
Duration of the post--AGB stage is relatively short and reduces with increasing initial
mass from $\approx 3.4\times 10^4$ yr for $M_0=1M_\odot$ to $\approx 600$ yr for
$M_0=4M_\odot$ (Miller Bertolami 2016).
Nuclear hydrogen shell burning is the dominant energy source in post--AGB stars and
the mass of the hydrogen envelope continues to decrease.
In most cases after the end of hydrogen shell burning the star becomes the white dwarf.
However there is a possibility that before the nuclear energy sources of the
star are exhausted a final thermal flash in the helium shell can occur
(Sch\"onberner 1979, 1983; Iben et al. 1983).
The final helium flash might take place on the stage of hydrogen shell burning (LTP)
or after the end of hydrogen burning (VLTP) on the white dwarf stage
(Bl\"ocker 2001).
In the HR diagram the evolutionary track the star experiences the loop extending to
the red giant domain with effective temperatures as low as $\approx 4000$\:K
(Althaus et al. 2005; Miller Bertolami et al. 2006).

Assumption on the final helium flash allows us to explain photometric and spectral variations
observed in some central stars of planetary nebulae:
FG~Sge (Sch\"onberner 1983; Iben 1984), V4334~Sgr (Duerbeck and Benetti 1996)
and V605~Aql (Clayton and De Marco 1997).
The time scale of such changes evaluated from stellar evolution computation is in a good
agreement with observations
(Bl\"ocker and Sch\"onberner 1997; Lawlor and MacDonald 2003; Miller Bertolami et al. 2006;
Miller Bertolami and Althaus 2007).

Among observed objects undergoing the final helium flash of most interest
is the variable FG~Sge which is a central star of the planetary nebula He 1--5
(Henize 1961).
Observational history of this star encompasses nearly 130 yr and during this interval
its effective temperature decreased from $\teff\approx 4.5\times 10^4$ K in 1880
to $\teff\approx 4500$ K in 1992 (Herbig and Boyarchuk 1968; van Genderen and Gautschy 1995;
Jeffery and Sch\"onberner 2006).
Moreover, during 60 yr the period of FG~Sge pulsations increased
from $\Pi\approx 5$ day in 1930 (van Genderen and Gautschy 1995) to $\Pi\approx 115$ day
in 1986--1989 (Arkhipova 1996).
Increase of the period of light variations of FG~Sge ceased after 1990
(Arkhipova et al. 1998, 2009).

The pulsation period $\Pi$ and the rate of period change $\dot\Pi$ are the most reliable
indicators of structural changes in the evolving star so that for more strict comparison
of the stellar evolution theory with observations one has to employ the methods of the stellar
pulsation theory.
For stars undergoing the final helium flash such works have not been done yet.
The present study is based on consistent stellar evolution and stellar pulsation
calculations.
In the framework of this approach the selected models of evolutionary sequences are used
as initial conditions for solution of the equations of radiative hydrodynamics and
time dependent convection describing radial stellar oscillations.
Dependencies of the pulsation period as a function of stellar evolution time $\Pi(\tev)$
obtained from computations are compared with observations of the pulsating variable FG~Sge.

\section*{results of stellar evolution calculations}

Stellar evolution calculations were done using the program MESA version 10398
(Paxton et al. 2018) with initial relative mass fractions of hydrogen and helium
$X_0=0.7$ and $Y_0=0.28$, respectively.
Initial abundances of elements heavier than helium were determined following
Grevesse and Sauval (1998).
The rates of nuclear energy generation and changes of element abundances
were calculated from solution of the kinetic equations for 29 isotopes from
hydrogen ${}^1\mathrm{H}$ to aluminium ${}^{27}\mathrm{Al}$ coupled by 51 reactions.
To this end we used the reaction network 'sagb\_NeNa\_MgAl.net' of the program MESA.
The reaction rates were calculated using the data base JINA Reaclib 
(Cyburt et al. 2010).

Convection was treated in the framework of the standard mixing length theory
(B\"ohm--Vitense 1958) with mixing length to pressure scale height ratio
$\alpha_\mathrm{MLT} = \ell/H_\mathrm{P} = 1.8$.
Additional mixing beyond the boundaries of convection zones due to overshooting
was calculated following the method proposed by Herwig (2000):
\begin{equation}
D_\mathrm{ov}(z) = D_0 \exp\left(-\frac{2z}{f H_\mathrm{P}}\right) ,
\end{equation}
where $D_0$ is the convection diffusion coefficient (Langer et al. 1985) in the layer
located in the convection zone on the distance of $0.004 H_\mathrm{P}$
from the Schwarzschild boundary,
$z$ is the radial distance from the boundary of the convection zone,
$f$ is the overshooting parameter.
To evaluate the role of overshooting in appearance of the final helium flash
we computed two evolutionary sequences with different values of $f$ for each
initial mass $M_0$.
In the first case the evolutionary calculations were done with the commonly used
value $f=0.016$ (Herwig, 2000), whereas the second evolutionary sequence was computed
in assumption of more efficient overshooting ($f=0.018$).

The mass loss rate $\dot M$ during the stage preceding AGB was computed by the
Reimers (1975) formula
\begin{equation}
\label{mdotr}
 \dot M_\mathrm{R} = 4\times 10^{-13} \eta_\mathrm{R} (L/L_\odot) (R/R_\odot) (M/M_\odot)^{-1} ,
\end{equation}
whereas on the AGB stage we employed the formula by Bl\"ocker (1995):
\begin{equation}
\label{mdotb}
 \dot M_\mathrm{B} = 4.83\times 10^{-9} \etab \dot M_\mathrm{R} (L/L_\odot)^{2.7} (M/M_\odot)^{-2.1} .
\end{equation}
All evolutionary sequences were computed with mass loss parameters
$\eta_\mathrm{R} = 0.5$ and $\etab = 0.05$.
Sufficiently reliable formulae for mass loss from post--AGB stars do not exist
so that to evaluate $\dot M$ on this evolutionary stage we used the Blocker formula
(\ref{mdotb}).
The parameter corresponding to the post--AGB stage is designated as $\etab^*$.
In order to evaluate the role of uncertainties in mass loss rates
the evolutionary calculations of the post--AGB stage were done with several values
of the mass loss parameter ranging within $0.02\le \etab^*\le 0.1$.
For the criterion of the post--AGB stage we used the condition
$M_\mathrm{env}/M \le 0.01$ (Miller Bertolami, 2016), where $M_\mathrm{env}$
is the mass of the hydrogen envelope and $M$ is the stellar mass.

\begin{figure}
\centerline{\includegraphics[width=14cm]{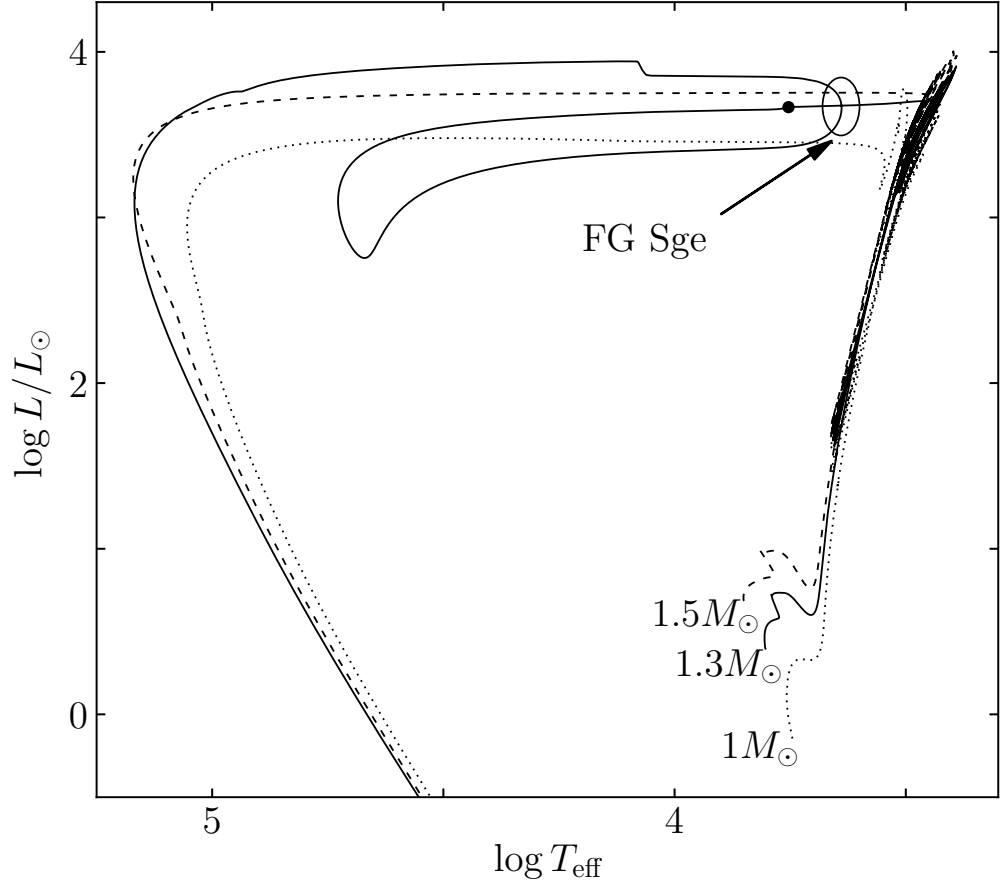}}
\caption{Evolutionary tracks of stars with initial mass $M_0=1M_\odot$ (dotted line),
         $1.3M_\odot$ (solid line) and $1.5M_\odot$ (dashed line) computed with
         parameters $f=0.016$ and $\etab^*=0.05$.
         The filled circle indicates the maximum energy release rate by the
         helium burning shell source.
         The oval in the upper right part of the figure marks the assumed location
         of FG~Sge in the HR diagram.}
\label{fig1}
\end{figure}

In the present study the stellar evolution calculations were done for stars with
initial masses $1M_\odot\le M_0\le 1.5M_\odot$ from the main sequence to the
white dwarf with luminosity as low as $L\sim 10^{-3}L_\odot$.
Evolutionary tracks were computed with initial mass step $\Delta M_0=0.02M_\odot$.
Results of computations are illustrated in Fig.~\ref{fig1} for stars with initial masses
$M_0=1M_\odot$, $1.3M_\odot$ and $1.5M_\odot$.
The overshooting parameter and the post--AGB mass loss parameter are
$f=0.016$ and $\etab^*=0.05$, respectively.
The loop in  the evolutionary track $M_0=1.3M_\odot$ on the post--AGB stage
is due to the final helium flash.
The maximum of energy release by the helium shell source is indicated by the
filled circle on the evolutionary track.
The oval shows the assumed location of FG~Sge in the HR diagram.

\begin{figure}
\centerline{\includegraphics[width=14cm]{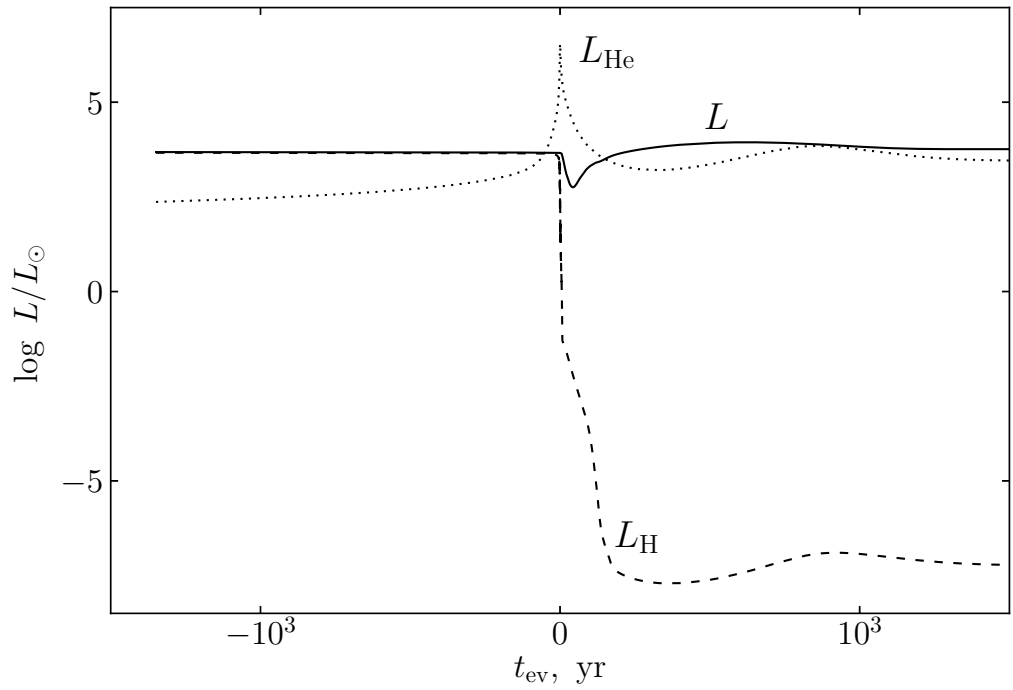}}
\caption{Time variations of the stellar surface luminosity $L$ (solid line) and
         the rates of energy release by the hydrogen $L_\mathrm{H}$ (dashed line)
         and the helium $L_\mathrm{He}$ (dotted line) shell sources during the final
         helium flash in the evolutionary sequence $M_0=1.3M_\odot$, $f=0.016$, $\etab^*=0.05$.}
\label{fig2}
\end{figure}

Variations with time $\tev$ of the surface luminosity $L$ and the rates of energy release
by the hydrogen $L_\mathrm{H}$ and helium $L_\mathrm{He}$ shell sources during
the helium flash are sfown in Fig.~\ref{fig2} for the evolutionary sequence $M_0=1.3M_\odot$,
$f=0.016$, $\etab^*=0.05$.
For the sake of convenience the time $\tev$ is set to zero at maximum $L_\mathrm{He}$.
As is seen, the nuclear burning in the hydrogen shell source ceases due to
abrupt increase of the energy release in the helium shell source.
During further evolution the helium--burning luminosity gradually decreases and finally
the star becomes the white dwarf.
Therefore, this is the final helium flash LTP.

\begin{figure}
\centerline{\includegraphics[width=14cm]{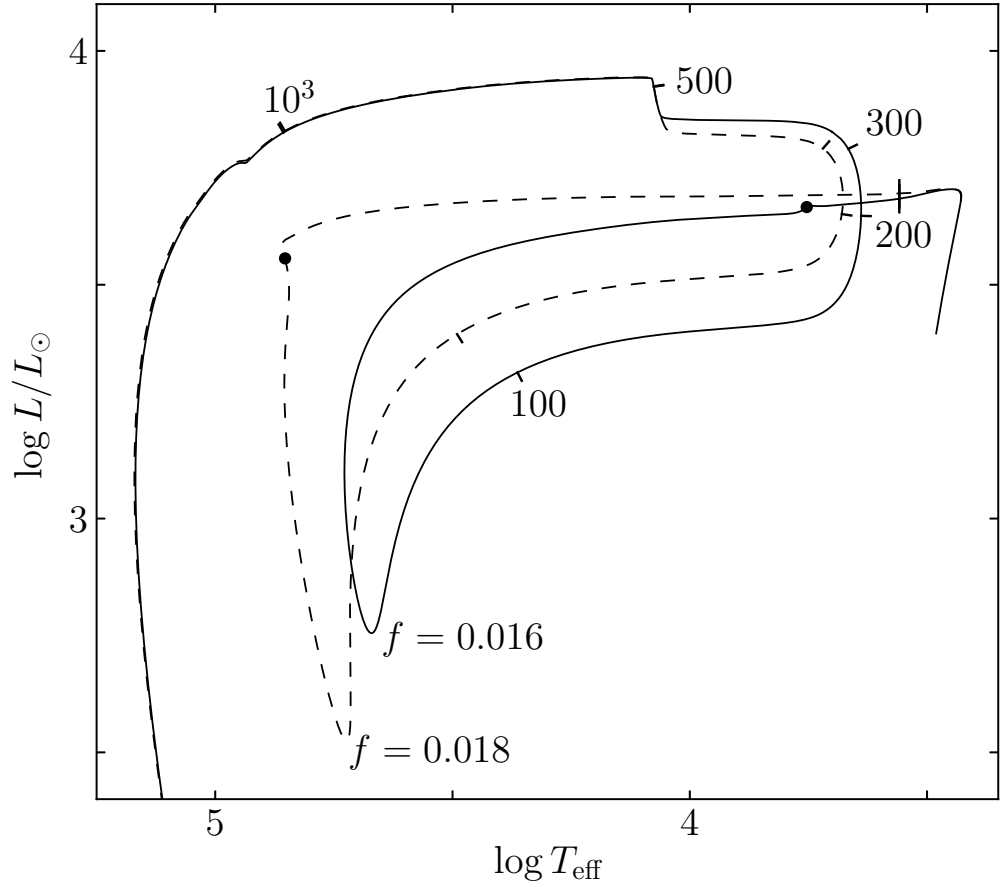}}
\caption{Evolutionary tracks of post--AGB stars with initial mass $M_0=1.3M_\odot$
         computed with overshooting parameters $f=0.016$ (solid line) and
         $f=0.018$ (dashed line).
         The parameter of the post--AGB mass loss rate is $\etab^*=0.05$.
         The vertical dash and the filled circle on the tracks mark the beginning
         of the post--AGB stage and the maximum rate of energy release by
         the helium shell source.
         Numbers near the curves indicate the time in years elapsed since the beginning
         of the post--AGB stage.}
\label{fig3}
\end{figure}

The role of overshooting in appearence of the final helium flash is illustrated
in Fig.~\ref{fig3},
where two evolutionary post--AGB tracks with initial mass $M_0=1.3M_\odot$ are plotted
for overshooting parameters $f=0.016$ and $f=0.018$.
Both tracks were computed with the post--AGB mass loss rate parameter $\etab^*=0.05$.
The vertical dash on the tracks indicates the time when the ratio of the hydrogen
envelope mass to the stellar mass is $M_\mathrm{env}/M = 0.01$.
As is seen, increase of the parameter $f$ leads to the later maximum of $L_\mathrm{He}$.
For example, the time interval between beginning of the post--AGB stage and the maximum
of $L_\mathrm{He}$ is 1350 yr for $f=0.016$, whereas for $f=0.018$ this time interval
increases to 4640 yr.
Nevertheless, the final helium flash is also LTP.
Moreover, notwithstanding the substantial initial difference in location of the tracks in
the HR diagram the stellar evolution after the maximum of $L_\mathrm{He}$
proceeds nearly in the same time scale and for $\tev \ge 500$ yr both tracks almost coincide.
Summarising the role of overshooting one has to note that for $f=0.016$
the final helium flash occurs at initial masses $1.30M_\odot\le M\le 1.32M_\odot$,
whereas for $f=0.018$ the corresponding initial mass interval becomes
$1.28M_\odot\le M\le 1.30M_\odot$.

\begin{figure}
\centerline{\includegraphics[width=14cm]{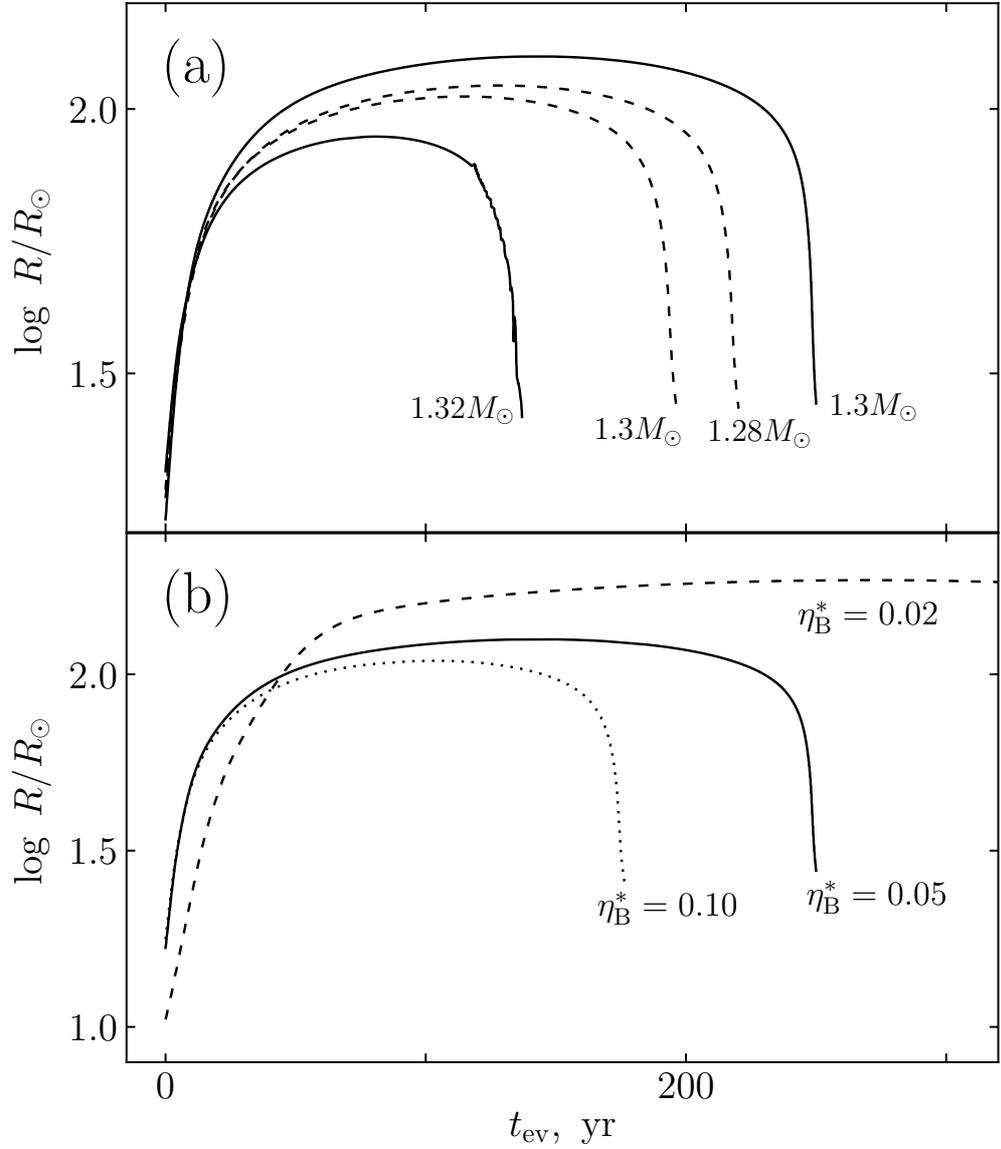}}
\caption{Time variations of the stellar radius at effective temperatures
         $\teff < 10^4$\:K.
         (a) -- Evolutionary sequences computed with overshooting parameter
         $f=0.016$ (solid lines) and $f=0.018$ (dashed lines).
         The numbers near the curves indicate the initial mass $M_0$.
         The post--AGB mass loss rate parameter is $\etab^*=0.05$.
         (b) -- Evolutionary sequences $M_0=1.3M_\odot$, $f=0.016$
         computed with mass loss rate parameters $\etab^*=0.02$, 0.05 and 0.1.}
\label{fig4}
\end{figure}

The period of radial pulsations and the stellar radius relate as $\Pi\propto R^{3/2}$,
therefore the time variation of the radius $R$ near the reddest point of the loop
in the evolutionary track allows us to evaluate the role of the initial mass $M_0$ as well as
parameters $f$ and $\etab^*$ in the rate of period change $\Pi(\tev)$ without
time consuming hydrodynamic computations.
Fig.~\ref{fig4}(a) shows the time variations $R(\tev)$ computed for $\etab^*=0.05$ in several
assumptions on the initial mass $M_0$ and the overshooting parameter $f$.
For the sake of convenience the time $\tev$ is set to zero when the decreasing effective
temperature reaches $\teff=10^4$\:K.
As is seen, increase of the initial mass from $M_0=1.3M_\odot$ to $M_0=1.32M_\odot$
for $f=0.016$ is accompanied by nearly twofold reduction of the time scale of radius
increase to its maximum value.
However in the case of more efficient overshooting ($f=0.018$) this time scale becomes
several times shorter.

The plots in Fig.~4(b) show the radius time variations for evolutionary sequences
$M_0=1.3M_\odot$, $f=0.016$ computed in three different assumptions on the
post--AGB mass loss rate: $\etab^*=0.02$, 0.05 and 0.1.
As is seen, decrease of the mass loss rate is accompanied by slower evolution near
the maximum radius although during initial stage of radius increase
the difference between the plots $\etab^*=0.05$ and $\etab^*=0.1$ is insignificant.

The parameters of evolutionary models at the maximum stellar radius during the LTP stage
are presented in Table~1.
Three first columns list the initial mass $M_0$,
the overshooting parameter $f$ and the mass loss parameter $\etab^*$.
In the fourth column we give the time $\tev$ measured from $\teff=10^4$\:K.
The following columns contain the mass $M$, the radius $R$, the luminosity $L$ and the effective
temperature $\teff$.

\section*{results of stellar pulsation calculations}

In the present study we carried out hydrodynamic computations of stellar pulsations for
stars on the LTP stage with effective temperatures $\teff < 10^4$\:K.
For initial conditions we used selected models of two evolutionary sequences
$M_0=1.3M_\odot$, $\etab^*=0.05$ computed with overshooting parameters $f=0.016$
and $f=0.018$.
The equations of radiation hydrodynamics and time--dependent convection for
radial stellar oscillations are discussed in our earlier papers
(Fadeyev, 2013; 2015).

\begin{figure}
\centerline{\includegraphics[width=14cm]{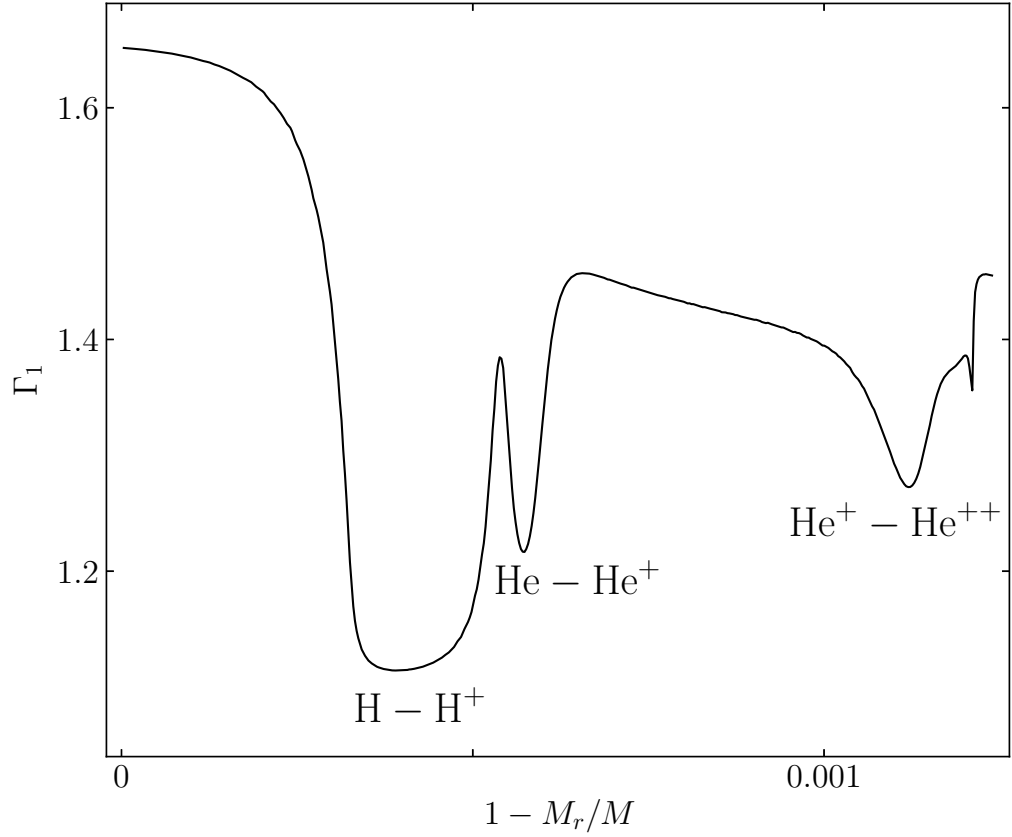}}
\caption{Adiabatic exponent $\Gamma_1$ as a function of mass coordinate
         in the initial conditions of the equations of hydrodynamics
         for the evolutionary sequence $M_0=1.3M_\odot$, $f=0.016$, $\etab^*=0.05$.
         The radius abd effective temperature are $R=125.4R_\odot$ and
         $\teff=4445$\:K, respectively.}
\label{fig5}
\end{figure}

Solution of the Cauchy problem for equations of hydrodynamics describes self--excited
stellar oscillations growing from the initial state of hydrostatic equilibrium.
All hydrodynamic models of stars on the LTP stage have the large rates of kinetic energy
growth: $\eta=\Pi d\ln E_\mathrm{K}/dt \sim 1$ due to the peculiar stellar structure.
Let us consider, for example, the model of the evolutionary sequence
$M_0=1.3M_\odot$, $f=0.016$, $\etab^*=0.05$ with effective temperature $\teff=4445$\:K.
The radius and the gas temperature at the inner boundary of the hydrodynamic model are
$r=5.5\times 10^{-3}R$ and $T=10^6$\:K, respectively.
Fig.~\ref{fig5} shows the plot of the adiabatic exponent $\Gamma_1=\left(\partial\ln P/\partial\ln\rho\right)_S$
as a function of the mass coordinate $q=1-M_r/M$, where $M_r$ is the mass of stellar matter
inside the radius $r$.
The strong instability against radial oscillations is due to substantial extension of the
hydrogen and helium ionization zones where the adiabatic exponent decreases below its
critical value: $\Gamma_1 < 4/3$.
Oscillations attain the large amplitude ($\delta r/r \sim 1$) because of strong pulsational
instability so that nonlinearity is the principal cause that pulsations are not strictly
periodic.

\begin{figure}
\centerline{\includegraphics[width=14cm]{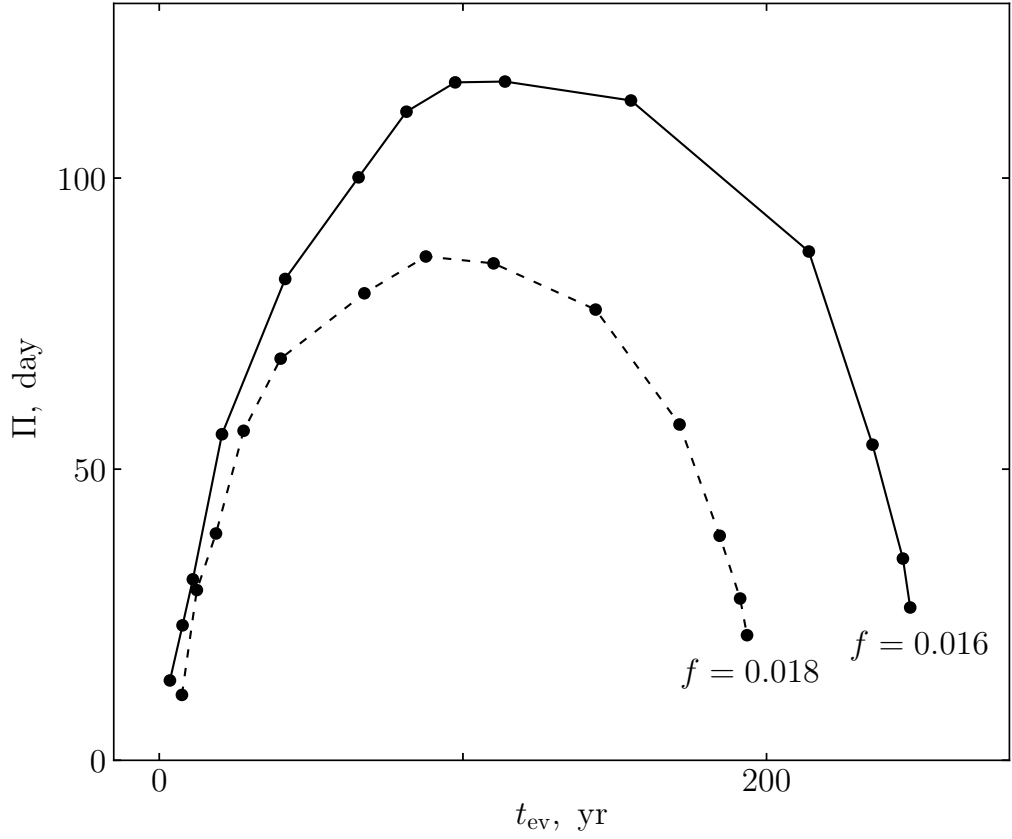}}
\caption{The period of radial oscillations $\Pi$ as a function of time
         $\tev$ for evolutionary sequences $M_0=1.3M_\odot$, $\etab^*=0.05$
         with overshooting parameter $f=0.016$ (solid line) and $f=0.018$
         (dashed line).
         Filled circles show periods evaluated using the hydrodynamic models.}
\label{fig6}
\end{figure}

As in  our preceding work on pulsating post--AGB stars (Fadeyev 2019) the period of radial
oscillations $\Pi$ was determined using the discrete Fourier transform of the kinetic energy
of pulsation motions $E_\mathrm{K}$.
Results of hydrodynamic computations are shown in Fig.~\ref{fig6} where time variations of the
period of radial oscillations are plotted for evolutionary sequences
$M_0=1.3M_\odot$, $\etab^*=0.05$ computed with overshooting parameters $f=0.016$ and $f=0.018$.
The maximum pulsation period of these dependences is $\Pi=117$ day for $f=0.016$
and $\Pi=87$ day for $f=0.018$.

To compare theoretical dependencies $\Pi(\tev)$ shown in Fig.~\ref{fig6} with observations
we used observational estimates of the period of light variations of FG~Sge presented by
Arkhipova and Taranova (1990), van Genderen and Gautschy (1995), Arkhipova (1996),
Arkhipova et al., (1998; 2009), Jurcsik and Montesinos (1999) and spanning
the time interval from 1960 to 2005 (see Fig.~7).
As is seen, increase of the period of FG~Sge can be fitted by the quadratic dependence
\begin{equation}
\label{fgsge}
 \Pi(t) = 19.87 + 4.439 (t-1960) -3.475\times 10^{-2} (t-1960)^2 ,
\end{equation}
where $1960 \le t \le 1990$.
The growth of the period ceased nearly in 1990 and the maximum period remaining roughly
constant for 15 years was $\Pi\approx 115$ day.
Unfortunately, determination of the mean value of the period of semiregular light variations of
long--period variable stars is accompanied by significant errors, so that at present
one cannot indicate the time corresponding to the maximum value of the period.

\begin{figure}
\centerline{\includegraphics[width=14cm]{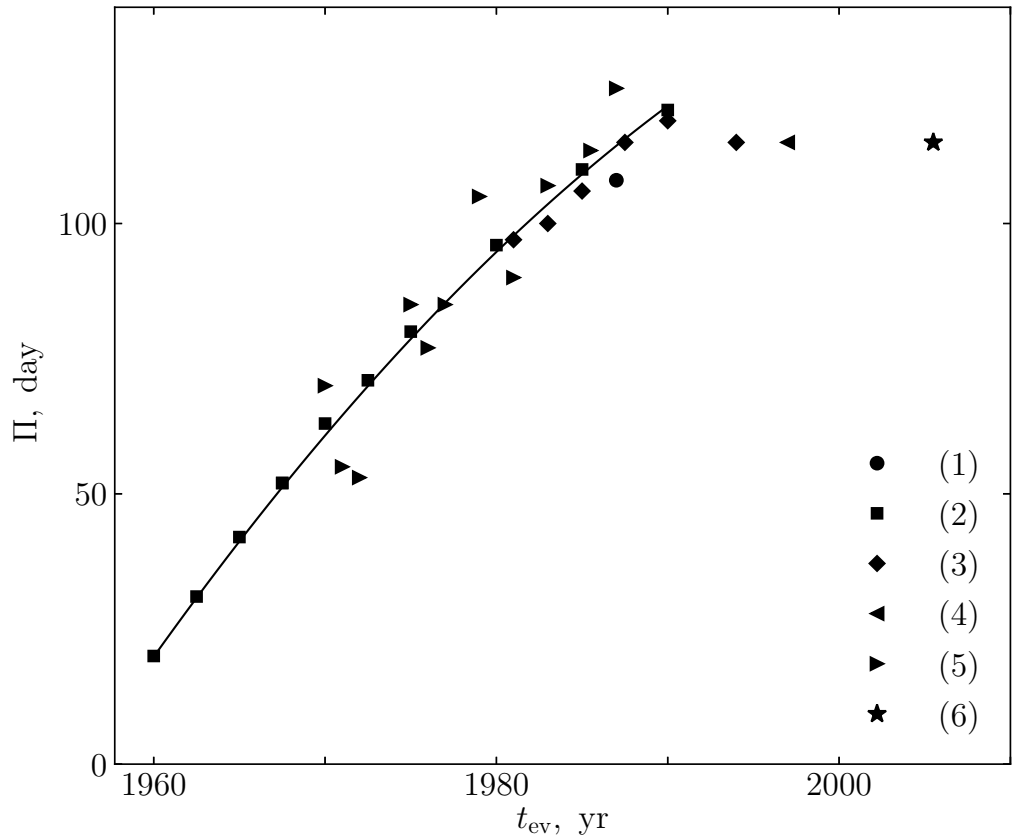}}
\caption{The period of light changes of FG~Sge from
         (1) Arkhipova and Taranova (1990);
         (2) van Genderen and Gautschy (1995);
         (3) Arkhipova (1996);
         (4) Arkhipova et al. (1998);
         (5) Jurcsik and Montesinos (1999);
         (6) Arkhipova et al. (2009).
         The solid line represents the quadratic fit (\ref{fgsge}) of the increasing period.}
\label{fig7}
\end{figure}

Comparison of the plots presented in Figs.~\ref{fig6} and \ref{fig7} allows us to conclude that
the evolutionary sequence $M_0=1.3M_\odot$, $\etab^*=0.05$, $f=0.016$ shows the better
agreement with observations.
In particular, the maximum pulsation period of this evolutionary sequence $\Pi=117$~day
is very close to obervational estimates of the period of FG~Sge after 1990.
At the same time one should note that between 1960 and 1990 the period of FG~Sge increased
with mean rate $\langle\dot\Pi\rangle \approx 3.3$~day/yr
which is more than three times larger than the theoretical mean period change rate
$\langle\dot\Pi\rangle \approx 0.96$~day/yr.

\section*{conclusions}

Results of stellar evolution calculations presented in this paper allow us to conclude
that in population~I stars with initial masses $1M_\odot\le M_0\le 1.5M_\odot$
the final helium flash LTP takes place within the narrow interval of initial masses $M_0$.
For mass loss rate parameters $\eta_\mathrm{R}=0.5$, $\eta_\mathrm{B}=0.05$
with overshooting $f=0.016$ this interval is $1.3M_\odot\le M_0\le 1.32M_\odot$.
In the case of more efficient overshooting ( $f=0.018$) the initial mass interval displaces
to somewhat smaller values: $1.28M_\odot\le M_0\le 1.3M_\odot$.
Variations of the post--AGB mass loss rate within a factor of five do not play a role in
occurence of the final helium flash but perceptibly affect the pulsation period change rate
on the LTP stage.
In particular, the time scale of evolution along the loop decreases with increasing
parameter $\etab^*$ whereas the reddest point of the loop corresponds to smaller values
of the radius and the pulsatuion period.

Consistent calculations of stellar evolution and nonlinear stellar pulsations
for the evolutionary sequence $M_0=1.3M_\odot$, $f=0.016$, $\etab^*=0.05$ show that
the maximum period of radial pulsations on the stage of LTP is $\Pi=117$.
This value is in a good agreement with observational estimates of the period of FG~Sge
obtained from 1990 to 2005 (Arkhipova, 1996; Arkhipova et al., 1998, 2009).
If we assume that FG~Sge is on the stage of its maximum radius then its mass, radius and
effective temperature are $M=0.565M_\odot$, $R=126R_\odot$ and $\teff=4445$\:K,
respectively.
However, as is seen in Fig.~\ref{fig6}, the change of the pulsation period near its maximum
within 10\% proceeds on the tile scale of $\sim 100$ yr.
Therefore, only reliable observational evidences in favour of FG Sge period decrease
in the near future could confirm our theoretical estimates of the mass and the radius.

The mean period change rate evaluated in the present study $\langle\dot\Pi\rangle = 0.96$ day/yr
is more than three times smaller than that of FG~Sge: $\langle\dot\Pi\rangle = 3.32$ day/yr.
Taking into account the growth of post--AGB evolution rate with increasing stellar
mass (Miller Bertolami, 2016; Fadeyev, 2019) one can suppose that better agreement between
the theory and observations could be obtained with somewhat higher mass of the post--AGB
star undergoing the final helium flash.
A possible solution of this problem could be found on the base of more extensive grid
of evolutionary tracks computed with various assumptions on the mass loss rates during
the stage of red giant before AGB as well as during the AGB stage.

\newpage
\section*{references}

\begin{enumerate}
\item L.G. Althaus, M.M. Miller Bertolami, A.H. C\'orsico, E. Garc\'ia--Berro, and P. Gil--Pons,
      Astron. Astrophys. \textbf{440}, L1 (2005).
\item V.P. Arkhipova, Astron. Lett. \textbf{22}, 743 (1996).
\item V.P. Arkhipova and O.G. Taranova, Astron. Lett. \textbf{16}, 347 (1990).
\item V.P. Arkhipova, V.F. Esipov and G.V. Sokol, Astron. Lett. \textbf{24}, 365 (1998).
\item V.P. Arkhipova, V.F. Esipov, N.P. Ikonnikova, G.V. Komissarova, and S.Yu. Shugarov,
      Astron. Lett. \textbf{35}, 534 (2009).
\item T. Bl\"ocker, Astron. Astrophys. \textbf{297}, 727 (1995).
\item T. Bl\"ocker and D. Sch\"onberner, Astron. Astrophys. \textbf{324}, 991 (1997).
\item T. Bl\"ocker, Astrophys. Space Sci. \textbf{275}, 1 (2001).
\item E. B\"ohm--Vitense, Zeitschrift f\"ur Astrophys. \textbf{46}, 108 (1958).
\item G.C. Clayton and O. De Marco, Astron. J. \textbf{114}, 2679 (1997).
\item R.H. Cyburt, A.M. Amthor, R. Ferguson, Z. Meisel, K. Smith, S. Warren, A. Heger,
      R.D. Hoffman, T. Rauscher, A. Sakharuk, H. Schatz, F.K. Thielemann, and M. Wiescher,
      Astrophys. J. Suppl. Ser. \textbf{189}, 240 (2010).
\item H.W. Duerbeck and S. Benetti, Astrophys. J. \textbf{468}, L111 (1996).
\item Yu.A. Fadeyev, Astron. Lett. 39, 306 (2013).
\item Yu.A. Fadeyev, MNRAS \textbf{449}, 1011 (2015).
\item Yu.A. Fadeyev, Astron. Lett. \textbf{45}, in press (2019).
\item N. Grevesse and A.J. Sauval, Space Sci. Rev. \textbf{85}, 161 (1998).
\item K.G. Henize, Publ. Astron. Soc. Pacific \textbf{73}, 159 (1961).
\item G.H. Herbig and A.A. Boyarchuk, Astrophys. J. \textbf{153}, 397 (1968).
\item F. Herwig, Astron. Astrophys. \textbf{360}, 952 (2000).
\item I. Iben, Astrophys. J. \textbf{277}, 333 (1984).
\item I. Iben, J.B. Kaler, J.W. Truran, J. W. and A. Renzini,  Astrophys. J. \textbf{264}, 605 (1983).
\item C.S. Jeffery and D. Sch\"onberner, Astron. Astrophys. \textbf{459}, 885 (2006).
\item J. Jurcsik and B. Montesinos, New Astron. Rev. \textbf{43}, 415 (1999).
\item N. Langer, M. El Eid and K.J. Fricke, Astron. Astrophys. \textbf{145}, 179 (1985).
\item T.M. Lawlor and J. MacDonald, Astrophys. J. \textbf{583}, 913 (2003).
\item M.M. Miller Bertolami, Astron. Astrophys. \textbf{588}, A25 (2016).
\item M.M. Miller Bertolami, L.G. Althaus, A.M. Serenelli and J.A. Panei,
      Astron. Astrophys. \textbf{449}, 313 (2006).
\item M.M. Miller Bertolami and L.G. Althaus, MNRAS \textbf{380}, 763 (2007).
\item B. Paczy\'nski, Acta Astron. \textbf{21}, 417 (1971).
\item B. Paxton, J. Schwab,  E.B. Bauer, L. Bildsten, S. Blinnikov, P. Duffell,
      R. Farmer,  J.A. Goldberg, P. Marchant, E. Sorokina, A. Thoul, R.H.D. Townsend, and F.X. Timmes,
      Astropys. J. Suppl. Ser. \textbf{234}, 34 (2018).
\item D. Reimers, \textit{Problems in stellar atmospheres and envelopes}
      (Ed. B. Baschek, W.H. Kegel, G. Traving, New York: Springer-Verlag, 1975), p. 229.
\item D. Sch\"onberner, Astron. Astrophys. \textbf{79}, 108 (1979).
\item D. Sch\"onberner, Astron. Astrophys. \textbf{272}, 708 (1983).
\item A.M. van Genderen and A. Gautschy, Astron. Astrophys. \textbf{294}, 453 (1995).
\end{enumerate}

\newpage
\begin{table}
\caption{Parameters of evolutionary models on the LTP evolutionary stage at the maximum stellar radius.}
\label{tabl1}
\begin{center}
 \begin{tabular}{cccccrrr}
  \hline
  $M_0/M_\odot$ & $f$ & $\etab^*$ & $\tev$, yr & $M/M_\odot$ & $\lg R/R_\odot$ & $\lg L/L_\odot$ & $T_\mathrm{eff}$, K \\
  \hline
 1.30 &  0.016  & 0.02  &   269 &   0.5667 &  2.267 &  3.857 &   3910 \\
      &         & 0.05  &   143 &   0.5647 &  2.099 &  3.744 &   4445 \\
      &         & 0.10  &   103 &   0.5645 &  2.039 &  3.696 &   4638 \\
 1.32 &         & 0.02  &   136 &   0.5662 &  2.029 &  3.783 &   4933 \\
      &         & 0.05  &    81 &   0.5661 &  1.948 &  3.726 &   5239 \\
      &         & 0.10  &    51 &   0.5661 &  1.865 &  3.689 &   5638 \\[4pt]
 1.28 &  0.018  & 0.02  &   195 &   0.5633 &  2.109 &  3.796 &   4532 \\
      &         & 0.05  &   128 &   0.5631 &  2.044 &  3.729 &   4696 \\
      &         & 0.10  &    96 &   0.5630 &  1.999 &  3.688 &   4832 \\
 1.30 &         & 0.02  &   176 &   0.5651 &  2.088 &  3.794 &   4638 \\
      &         & 0.05  &   114 &   0.5650 &  2.024 &  3.730 &   4812 \\
      &         & 0.10  &    84 &   0.5649 &  1.974 &  3.690 &   4980 \\
  \hline          
 \end{tabular}
\end{center}
\end{table}
\clearpage

\newpage
\section*{figure captions}

\begin{itemize}
 \item[Fig. 1.] Evolutionary tracks of stars with initial mass $M_0=1M_\odot$ (dotted line),
                $1.3M_\odot$ (solid line) and $1.5M_\odot$ (dashed line) computed with
                parameters $f=0.016$ and $\etab^*=0.05$.
                The filled circle indicates the maximum energy release rate by the
                helium burning shell source.
                The oval in the upper right part of the figure marks the assumed location
                of FG~Sge in the HR diagram.

 \item[Fig. 2.] Time variations of the stellar surface luminosity $L$ (solid line) and
                the rates of energy release by the hydrogen $L_\mathrm{H}$ (dashed line)
                and the helium $L_\mathrm{He}$ (dotted line) shell sources during the final
                helium flash in the evolutionary sequence $M_0=1.3M_\odot$, $f=0.016$, $\etab^*=0.05$.

 \item[Fig. 3.] Evolutionary tracks of post--AGB stars with initial mass $M_0=1.3M_\odot$
                computed with overshooting parameters $f=0.016$ (solid line) and
                $f=0.018$ (dashed line).
                The parameter of the post--AGB mass loss rate is $\etab^*=0.05$.
                The vertical dash and the filled circle on the tracks mark the beginning
                of the post--AGB stage and the maximum rate of energy release by
                the helium shell source.
                Numbers near the curves indicate the time in years elapsed since the beginning
                of the post--AGB stage.

 \item[Fig. 4.] Time variations of the stellar radius at effective temperatures
                $\teff < 10^4$\:K.
                (a) -- Evolutionary sequences computed with overshooting parameter
                $f=0.016$ (solid lines) and $f=0.018$ (dashed lines).
                The numbers near the curves indicate the initial mass $M_0$.
                The post--AGB mass loss rate parameter is $\etab^*=0.05$.
                (b) -- Evolutionary sequences $M_0=1.3M_\odot$, $f=0.016$
                computed with mass loss rate parameters $\etab^*=0.02$, 0.05 and 0.1.
                
 \item[Fig. 5.] Adiabatic exponent $\Gamma_1$ as a function of mass coordinate
                in the initial conditions of the equations of hydrodynamics
                for the evolutionary sequence $M_0=1.3M_\odot$, $f=0.016$, $\etab^*=0.05$.
                The radius abd effective temperature are $R=125.4R_\odot$ and
                $\teff=4445$\:K, respectively.

 \item[Fig. 6.] The period of radial oscillations $\Pi$ as a function of time
                $\tev$ for evolutionary sequnces $M_0=1.3M_\odot$, $\etab^*=0.05$
                with overshooting parameter $f=0.016$ (solid line) and $f=0.018$
                (dashed line).
                Filled circles show periods evaluated using the hydrodynamic models.

 \item[Fig. 7.] The period of light changes of FG~Sge from
                (1) Arkhipova and Taranova (1990);
                (2) van Genderen and Gautschy (1995);
                (3) Arkhipova (1996);
                (4) Arkhipova et al. (1998);
                (5) Jurcsik and Montesinos (1999);
                (6) Arkhipova et al. (2009).
                The solid line represents the quadratic fit (\ref{fgsge}) of the increasing period.

\end{itemize}

\newpage
\clearpage

\newpage
\clearpage

\newpage
\clearpage

\newpage
\clearpage

\newpage
\clearpage

\newpage
\clearpage

\newpage
\clearpage

\end{document}